\documentclass[12pt, epsfig, preprint]{aastex}
\def\RSUN{\rm R$_{\odot}$}

\def\LA{Ly$\alpha$ }
\def\LB{Ly$\beta$ }

\def\RSUN{\rm R$_{\odot}$}

\begin{document}
\title{Transition Region Emission and Energy Input to Thermal
Plasma During the Impulsive Phase of Solar Flares}

\author{John C. Raymond}
\affil{Harvard-Smithsonian Center for Astrophysics, 60 Garden St., Cambridge, MA 02138}



\author{Gordon Holman}
\affil{NASA's GSFC, Code 612.1, Greenbelt, MD 20771}

\author{A. Ciaravella}
\affil{INAF-Osservatorio Astronomico di Palermo, P.za Parlamento 1, 90134, Palermo, Italy}
 
\author{A. Panasyuk, Y.-K. Ko, and J. Kohl }
\affil{Harvard-Smithsonian Center for Astrophysics, 60 Garden St., Cambridge, MA 02138}

 
\begin{abstract}
The energy released in a solar flare is partitioned between thermal
and non-thermal particle energy and lost to thermal conduction
and radiation over a broad range of wavelengths.  It is 
difficult to determine the conductive losses and the energy
radiated at transition region temperatures during the impulsive
phases of flares.  We use UVCS measurements of O VI
photons produced by 5 flares and subsequently scattered by O VI
ions in the corona to determine the 5.0 $\le$~\rm log T $\le$ 6.0 transition 
region luminosities.  We compare them with the rates of increase of thermal
energy and the conductive losses deduced from RHESSI and GOES 
X-ray data using areas from RHESSI images to estimate 
the loop volumes, cross-sectional areas and scale lengths. 
The transition region luminosities during the 
impulsive phase exceed the X-ray luminosities for the first few minutes,
but they are smaller than the rates of increase of thermal energy
unless the filling factor of the X-ray emitting gas is $\sim 0.01$.
The estimated conductive losses from the hot gas are too large to be
balanced by radiative losses or heating of evaporated plasma, and 
we conclude that the area of the flare magnetic flux
tubes is much smaller than the effective area measured by RHESSI during this phase
of the flares.  For the 2002 July 23 flare, the energy
deposited by non-thermal particles exceeds the X-ray and UV energy losses
and the rate of increase of the thermal energy.

\end{abstract}


\section{Introduction} 
 
The energy released in solar flares appears in a variety of forms, including 
electromagnetic radiation, energetic electrons and ions, heated plasma, and 
mass motions ranging from local turbulence to coronal mass ejections (CMEs).  
To understand the physical mechanisms responsible for this 
energy release and transport we need to know how energy is 
partitioned among these channels.  Therefore, evaluating the energy 
contained in these various components is an important part of solar flare 
research (e.\ g., Emslie et al.\ 2004).  

A significant fraction of the energy released in moderate to large flares 
appears as hot plasma with a temperature on the order of 10 to 40 million 
degrees.  X-ray observations of the bremsstrahlung spectrum of 
this plasma provide temperatures and emission measures which, along with 
estimates of the plasma volume, allow the energy content of the plasma to be 
computed.  In large flares the maximum value of this thermal energy content is 
typically found to be comparable to but less than the computed accumulated 
energy in nonthermal electrons (Holman et al.\ 2003, Emslie et al.\ 2004).  

Despite substantial progress in our ability to determine these energies, there 
are still significant uncertainties (e. g., Emslie et al.\ 2005).  For the 
thermal energy content, one of these is energy losses.  Based on numerical
simulations of flare loops, Isobe, Takasaki \& Shibata (2005) estimate 
radiative losses at 40\% of the energy deposited.  On the observational
side, Kane, Frost \& Donnelly (1979) found a close relationship between 
EUV and hard X-ray intensities during the impulsive phases of flares, and
 full-disk spectral observations of a few flares with the SORCE and 
TIMED satellites (Brekke et al. 1996; Woods et al. 2004) show EUV brightenings
that imply significant emission at transition region temperatures.  
RHESSI observations of the positron annihilation line indicate that much 
of the annihilation occurred at temperatures of 2 to $\rm 7 \times 10^5$ K 
and densities above $10^{13}~\rm cm^{-3}$ during the impulsive 
phase of at least one flare (Share et al. 2003, 2004), implying large
emission measures and radiation losses at transition region temperatures.  
To the extent that such energy losses are significant, the energy required to heat the plasma 
is greater than the computed instantaneous energy content.  

Bremsstrahlung losses are easily measured, but thermal conduction of 
heat into the transition region and related losses are more difficult 
to determine.  Conduction losses are often estimated with 
the assumption of classical conduction (Spitzer conductivity), taking the 
conduction scale length to be the flare loop half length (e. g., Veronig et 
al.\ 2005).  However, the conductivity may not be classical and this scale 
length may not be correct.  In determining the total energy devoted to heating 
plasma, it is also possible that cooler plasma not observable in X-rays is 
directly heated by energetic particles.  Therefore, the best approach would be to observe the flare 
plasma in all temperature regimes.  This is generally not possible, however.  

The development of a large solar flare typically occurs
in two parts; the impulsive and gradual phases.  The
former is attributed to the rapid acceleration of particles,
which then deposit their energy in the chromosphere and
produce hard X-rays.  The latter is associated with the 
softer, longer lasting X-ray emission from gas evaporated
into the corona.  Because the impulsive phase occurs in
a small region and only lasts a short time,  a
spectrograph making raster scans over an active region
is unlikely to catch the impulsive phase.  Therefore, most spectroscopic
observations of flares pertain to the gradual phase.  

In view of the above discussion, it is important to constrain the amount of 
plasma at transition region temperatures in the impulsive phase of flares in 
order to assess where energy is deposited and to constrain the thermal 
conduction losses from the hot flare plasma.  The amount of plasma at transition 
region temperatures is also related to the rate of evaporation of chromospheric 
material into the flare loops.  

In this paper we report measurements of the luminosities at transition region 
temperatures, $L_{TR}$, for several X-class flares during their impulsive phases.
The measurements rely on spectra from the Ultraviolet Coronagraph
Spectrometer (UVCS) of O VI photons from the flare scattered by 
O VI ions in the corona. Section 2 presents the UVCS observations
for flares observed on 2002 July 23 and August 24, and 2003 Oct. 23, 
Nov. 2 and Nov. 4.   We compare
the results with RHESSI and GOES X-ray observations.  Other aspects of the UVCS 
observations of the 2002 events are presented in Raymond et al.\ (2003),
and RHESSI observations of the 23 July 2002 event are presented in a 
special issue of ApJ Letters (2003, ApJ Letters 595).  Section 3 
presents the analysis of the UVCS observations, section 4 describes the analysis of the X-ray observations, and 
section 5 summarizes the observed luminosities and compares
the radiative losses, thermal conduction losses, energy deposition by
non-thermal particles and the rate of change of thermal energy in the
X-ray emitting plasma.

\section{UVCS Observations}

UVCS (Kohl et al. 1995) is a longslit UV spectrograph fed by a
coronagraph.  The 40$^\prime$ slit is placed parallel to a tangent
to the solar limb at heliocentric distances between 1.5 and 10 \RSUN .
In all the cases discussed here, UVCS was performing a CME watch,
taking a long series of short exposures above an active region near
the limb that had been chosen for its likelihood to produce a Coronal Mass Ejection (CME).
The observations presented here consist of sequences of 120 second
exposures separated by about 10 second readout intervals.  The
UVCS slit was placed at 1.63 \RSUN , and the 100 $\mu$ slit width
provided spectral resolution of about 0.4 \AA .  In all cases
one panel of the detector mask covered the 1024 - 1042 \AA\/ range,
and other panels covered lines of ions such as C III, Si XII
and [Fe XVIII].  Here we will discuss mainly the O VI lines.  The
line intensities were determined based on current calibration files,
including the loss of sensitivity over time for narrow settings
of the internal occulter based on Gardner et al. (2002).

In all five cases a CME occurred, but during the interval between the
flare onset in hard X-rays and the time the CME reached the UVCS slit,
the lines of O VI at 1031.9\AA\/ and
1037.6\AA\/ brightened.  In all cases it is obvious when the CME reaches the
UVCS slit because of large disturbances in the Doppler widths and
velocity centroids of the lines, and because O VI and other lines
fade rapidly as hot, low density plasma displaces the pre-CME
streamer gas.  

We interpret the O VI brightening before the CME reaches the UVCS slit as radiation from the flare 
scattered by O VI ions in the corona.  We are confident of this interpretation for 
several reasons:  1) The O VI brightening 
occurs simultaneously along the entire length of the slit, while
the CME disturbance affects the center of the slit first, and the
disturbance spreads along the slit over the course of several
exposures (see Figure 2 of Raymond et al. 2003).  2) The additional O VI emission shows an
intensity ratio I(1032)/I(1037) of approximately 4:1 (Figure 1, right panel), characteristic of scattering
of solar disk O VI lines, rather than the 2:1 ratio
produced by collisional excitation of O VI ions in the corona.
3) The centroids and widths of the line profiles are unchanged from their pre-CME 
values. 4) The O VI emission peaks coincide with the RHESSI hard
X-ray peaks in the three events where the comparison can be made
(Figures 1 and 2).

Figure 1 shows the intensities of the O VI 1032\AA\/ line and the RHESSI hard X-ray and
GOES soft X-ray fluxes for the 2003, Nov. 4 event. The other four
events are shown in Figure 2. 
We have summed the O VI flux along the entire
UVCS slit after subtracting as background the average
of about 50 pre-flare exposures.  A uniform background on the UVCS detector can
arise during the flare due to X-rays penetrating the front
of the UVCS instrument or to energetic particles.  The contributions
of X-rays and energetic particles can be identified by the correlation of the background with GOES
X-ray and proton fluxes, respectively, but for the present purposes it only matters
that the background is very smooth on \AA ngstrom scales.  We therefore
integrate over a 1 \AA\/ band centered on the O VI $\lambda$1032
line and subtract as background a neighboring band centered 1.2 \AA\/
toward shorter wavelengths.  For most events, the rise in background
only occurs after the O VI spike we discuss here.  The X-ray light curves
are the GOES energy fluxes in the 0.5 to 4 and the 1 to 8 \AA\/ bands,
along with the RHESSI count rates in the 50 to 100 keV band.  The RHESSI observations
did not cover the second O VI peak of the 2003, Nov. 4 event or the impulsive peaks
of the 2002, Aug. 24 and 2003, Oct. 23 events.

For more detailed analysis we measured the intensities of the
1032 and 1037\AA\/ lines in portions of the UVCS slit large enough
to provide good signal to noise, typically 8 segments.  The intensities
were extracted by fitting Gaussians plus continuum.  Here we consider
for each exposure only segments of the slit in which the line widths
and shifts were unchanged from the pre-CME values.  
For three events we used the entire slit when possible,
but after the CME disturbs the central portion we use the
ends of the slit, which remain undisturbed for another 1 or 2 exposures.
As seen in Figure 2 of Raymond et al. (2003), the O VI intensity near
the middle of the slit drops as the CME blows away the ambient plasma,
and some brightening is seen at the edges of the CME bubble. In
all five events, Figures 1 and 2 show a drop in O VI intensity just
after the intensity spike, because the CME removes most of the coronal
plasma capable of producing or scattering O VI~photons.
Table 1 lists the parameters of the observations.

\section{Analysis of the UV Observations}

Several steps are required to obtain the transition region luminosity of an impulsive
flare.  First, we determine the O VI column density in the pre-flare
corona.  Then, we use that column density and  the scattering cross section
to determine the illuminating flux due to the flare.  Finally, we use the ratio of the
total radiative loss rate to the O VI emissivity to find the transition
region cooling.

{\it Pre-CME O VI column density:}  The observed O VI intensities originate
from both collisional excitation in the corona and resonance scattering
of O VI photons from the solar disk (Mariska, 1977; Noci, Withbroe \& Kohl 1987).  The collisional
component has an intensity ratio I(1032)/I(1037) = 2:1, while the radiative
component has a ratio of 4:1.  Simple algebra yields the
ratio of collisional to radiative components, and it equals

\begin{equation}
{\frac{I_{coll}(1032)}{I_{rad}(1032)}} ~ = ~ \frac{q_{1032} n_e n_{O VI}}{n_{O VI} \sigma_{eff} W I_{disk}(1032)} ,
\end{equation}

\noindent
where $q_{1032}$ is the collisional excitation rate taken from CHIANTI (Young et al. 2003), 
$n_e$ and $n_{O VI}$ are the electron and $O^{+5}$ densities, $\sigma_{eff}$ is the effective 
scattering cross section obtained by convolving the scattering profile and the solar O VI emission profile, and 
W is the dilution factor averaged along the line of sight.  W depends on the distribution of plasma
along the line of sight, and we assume a spherically symmetric corona with density declining as
$r^{-4}$ (e.g., Gibson et al. 1999).  We assume a coronal line width $v_{1/e}$ = 60 $\rm km~s^{-1}$
(Kohl et al. 1997) and a solar disk emission profile width of 30 $\rm km~s^{-1}$.
The disk O VI intensity measured by UVCS (Raymond et al. 1997), $I_{disk}(1032)$, was multiplied by a factor
of 2 to account for the increased brightness near solar maximum and the proximity of the
slit to a bright active region (Woods et al. 2005; Ko et al. 2002).

Equation 1 provides a value for $n_e$ (e.g, Ko et al. 2002), since $n_{O VI}$ cancels out.
We then derive the O VI column density, $N_{O VI}$, from

\begin{equation}
N_{O VI} ~=~ \frac{4 \pi I_{coll}(1032) }{q_{1032} n_e}~~~~~ cm^{-2} .
\end{equation}

\noindent
{\it Flare O VI luminosity:} Given the O VI column density, we can derive the 
luminosity of the flare from the additional
O VI intensity observed during the impulsive phase

\begin{equation}
L_{flare}(1032)~=~ {\frac{4 \pi I_{flare}(1032)}{N_{O VI} \sigma_{eff}^\prime W^\prime }} ~~~~ photons~s^{-1}  ,
\end{equation}

\noindent
where $I_{flare}(1032)$ is the 1032 intensity spike observed by UVCS due to the flare.
Note that the dilution factor $W^\prime $ =  $< 1/(4 \pi r^\prime$$ ^2) >$, where $r^\prime$ is the distance
from the flare to a point in the UVCS field of view.  It differs from W in Equation 1, because 
the flare luminosity originates effectively in a point source, while the pre-flare illumination arises from 
the solar disk.  For $W^\prime$ we assume, as for W, a $1/r^4$ spherical density distribution and compute
the weighted average along the line of sight.

An interesting complication is that the effective scattering cross section, $\sigma_{eff}^\prime$, 
can be affected by Doppler dimming if the flare O VI emission
profile is Doppler shifted.  High velocity jets, sprays or CME material
will be Doppler shifted well away from the coronal absorption lines (e.g. Pike \& Mason 2002; Innes
et al 2001), so the UVCS O VI enhancements are insensitive to any emission from that gas.
Instead, the UVCS enhancement is determined by plasma at transition region temperatures 
at the base of the flaring loop system.  Doschek, Feldman \& Rosenberg (1977) observed blue-shifts
of 50-80 $\rm km~s^{-1}$ in transition region lines of Si IV, C IV and N V during the impulsive
phase of a flare.  Czaykowska et al. (1999) observed a brief downflow of about 20 $\rm km~s^{-1}$
followed by an evaporative upflow of 50 $\rm km~s^{-1}$ in O V $\lambda$630 in the gradual phase of a flare on 1998 Apr. 29.
Milligan et al. (2006) observed a 43 $\rm km~s^{-1}$ downflow in O V during an explosive M-class flare.
X-ray observations sometimes show upflows of several hundred $\rm km~s^{-1}$ (e.g.
Doschek, Mariska \& Sakao 1996) during the impulsive phase, and constant mass flux would suggest
upflows of order 50 $\rm km~s^{-1}$ at transition region temperatures.  

We can constrain the Doppler
dimming effect, because the upflow should be directed toward the portion of the UVCS slit
directly above the flare, while the ends of the UVCS slit will see only the component of the
upflow velocity in that direction, or about 50\% of the upflow speed.  The requirement that
the flare luminosity be the same for all portions of the slit then constrains the upflow speed.  
Figure 3 shows the derived values of the O VI luminosity as a function of position along the
UVCS slit for one exposure during the 2003 Nov. 2 event for several choices of upflow speed.  
The luminosity must, of course, be the same all along the slit, but upflow speeds 
below 80 $\rm km~s^{-1}$ leave a dip near the
middle of the slit, while a speed of 100 km/s creates a hump.  We conclude that the upflow
speed is about 90 $\rm km~s^{-1}$.   We use that value for all the exposures of this event,
because the first two exposures were too noisy for a reliable determination, and in the last
exposure the CME had disrupted the center of the slit, forcing us to use only the 4 bins at
the ends of the slit.  We find upflow speeds of 50 $\rm km~s^{-1}$ for the 2002, Aug. 24 and
2003, Oct. 23 events, and 90 $\rm km~s^{-1}$ for the 2002 July 23 and 2003 November events.  
From Figure 3, it can be seen that the choice of upflow speed affects the derived O VI 
luminosities by about a factor of 1.5 level when integrated along the slit.

{\it Transition Region luminosity:} We determine the emission measure (EM) of the O VI emitting gas
by dividing the O VI luminosity by the emissivity from CHIANTI using the Mazzotta et al. (1998)
ionization balance and Grevesse \& Sauval (1998) abundances.  This EM is then
multiplied by a total cooling rate (Raymond, Cox \& Smith 1976) modified for
densities of $10^{12}~\rm cm^{-3}$ and  the
solar photospheric abundances of Grevesse \& Sauval (1998) to obtain the power 
radiated from temperatures near log T = 5.5. To estimate the cooling from the transition region
as a whole, we assume that the Emission Measure EM($log T$) scales with 
temperature as appropriate for a thermal conduction boundary layer in 
equilibrium or an evaporative flow (Raymond \& Doyle 1981), and we then 
multiply by the cooling rates.  For this study 
we define the Transition Region to be 5.0 $\le$~\rm log T $\le$ 6.0.

Table 2 shows the derived O VI $\lambda$1032 luminosities and emission measures $EM_{UV}$
(in units of $10^{48}~\rm cm^{-3}$) and the inferred power radiated at transition
region temperatures, $L_{TR}$, along with the GOES X-ray 
temperatures $T_6$ and emission measures $EM_X$ and the effective areas of the
X-ray emission from RHESSI images in units of $10^{18}~\rm cm^2$ (see section 4).  
We note that the uncertainties in the luminosities derived from the ends of the slit after the center 
of the slit has been disturbed by the CME should be larger than the others.  However, the decline in  O VI
luminosity for the July 23 event shown in Table 2 is similar to the decline in C III stray 
light brightness shown in Figure 3 of Raymond et al. (2003), suggesting that the trend is
correct.   We also note that the
Asplund, Grevesse \& Sauval (2004) set of abundances have less oxygen and would require Emission
Measures larger by 50\%, and the transition region luminosity, $L_{TR}$, would increase by about 25\%.
Overall, considering the uncertainties in $N_{OVI}$, the Doppler dimming, and the
dependence of the dilution factor on the distribution of O VI along the line of sight, we
estimate that the values of $L_{TR}$ in Table 2 are good to better than a factor of 2.

\section{X-ray Observations}

X-ray observations were obtained from the GOES and RHESSI satellites.  We used the GOES
data to characterize the thermal X-ray emission and the RHESSI data to determine the 
non-thermal emission and the size of the emitting region.

For all the flares, we measured the effective area of the X-ray
source region, which we took to be the area contained within the 50\% of peak flux contour
in the 12-25 keV band.
During the 2002 Aug. 24 flare, RHESSI went into orbital night shortly after 00:57:00 UT,
and during the 2003 October 23 flare it was only possible to obtain an image during the last
UVCS exposure.  For the 2003 Nov. 2 and Nov. 4 flares, we only have images for the first 2
and 3 UVCS exposures, respectively.  
For the 2002 July 23 flare, Holman et al.\ (2003) obtained thick-target fits 
of double-power-law electron distributions with a low-energy cutoff to the spectra above the 
isothermal component.  The minimum nonthermal electron energy flux, $L_{NT}$, for each time interval was 
derived from those fits for that flare. 

Temperatures, emission measures and total radiative losses of the thermal X-ray emitting plasma
were derived from the GOES X-ray fluxes using the GOES Workbench in the
Solar Software Package (e.\ g., D.M. Zarro and K. Tolbert 2006, \break
http://orpheus.nascom.nasa.gov/~zarro/idl/goes/goes.html).  We note that the recent results of White et al.\ (2005)
using coronal abundances are incorporated into the software, in particular the effects of
assuming ``coronal" rather than photospheric abundances and CHIANTI (Young et al. 2003) atomic rates.
The derived emission measures are therefore smaller and the total radiative losses larger than 
would have been derived using earlier software.  For example, the emission measures 
for the July 23 event are about half those shown in Holman et al. (2003).  Temperatures 
and emission measures could not be derived for the last two UVCS exposures for the 2003 Nov. 4
flare, because the GOES data were saturated.  
Table~3 shows the total luminosity of the 
the X-ray emitting gas, $L_{therm}$, computed from the temperatures and emission measures from Table 2,
along with the total conductive power (see section 5).  Table~3 also includes the minimum
nonthermal energy deposition rate, $L_{NT}$, for the July 23 event from Holman et al.\ (2003) and the 
rate of change of thermal energy in the X-ray emitting gas, dE/dt.

We compute dE/dt from the thermal energy content computed from the GOES spectra and RHESSI
effective areas.
The energy content of the thermal plasma as a function of time $t$, written as a function of 
the temperature $T$ and emission measure $EM$, is

\begin{equation}
E_{th}(t) ~=~ 3kT(t)[EM(t) f V]^{1/2} ,
\end{equation}

\noindent
where $V$ is the volume of the plasma and $f \leq 1$ is the volume filling factor.  The maximum
thermal energy content is obtained for a filling factor of 1.  We estimate the values of $V$ from 
the areas $A$ of the X-ray source from Table~2 and calculate the volume to be $V = A^{3/2}$.  
The mean increase of the thermal energy was calculated for each 120~s UVCS exposure.  
The results are shown in the sixth column of Table~3.  Negative values correspond to periods when there was 
a net loss of thermal energy.  

We note that there are two possible contributions to dE/dt; simply heating coronal gas from
coronal to flare temperatures or heating chromospheric gas to flare temperatures in an
evaporative flow that passes through the transition region temperature range.  In
most models, the latter dominates, in accordance with the high densities observed in 
flares (e.g. Feldman, Doschek \& Kreplin 1982), though Feldman et al. (2004) attribute the
high densities to compression of the coronal plasma.

\section{Discussion}

From Table 3 (third and fourth columns) it is apparent that the transition region 
luminosities are comparable to the X-ray luminosities in all cases.  In general, 
$L_{TR}$ is larger for the first few minutes. Then the X-ray flux becomes larger,
presumably as chromospheric material is evaporated into the corona.  However, as 
shown in Table 3, the transition region luminosity (third column) seldom exceeds 10\%
of the rate of change of thermal energy in the X-ray emitting plasma (column 6).  
Therefore, we find that heat conducted from the X-ray emitting plasma to lower 
temperature plasma and lost as radiation from plasma at 5.0 $\le$~\rm log T $\le$ 6.0 
transition region temperatures does not dominate the energetics of the X-ray 
emitting plasma.  

However, we have estimated the thermal energy of the flare from the X-ray temperatures and 
emission measures of the thermal components, along with the assumption that the 
volume approximately equals the emitting area $A$ seen in the RHESSI images times 
a thickness of $A^{1/2}$.  This estimate assumes a filling factor of 1.  A filling 
factor of 0.1 would reduce the thermal energies and their rates of change by a 
factor of 3. A filling factor of 0.01 would be required to make $L_{TR}$ comparable
to dE/dt.  The densities derived from EM/$A^{3/2}$ with f=1 range from a 
$10^{10}$ to a few times $10^{11}~\rm cm^{-3}$ so f=0.1 to 0.01 would not require an
unreasonably high density  compared with the densities of order $10^{12}~\rm cm^{-3}$
from the models of Isobe et al. (2005) or density measurements based on X-ray line
ratios (e.g. Doschek et al. 1981).  Therefore, the values for dE/dt
could be brought down to values comparable to the radiative losses
if the filling factors are as low as 0.01, as inferred by Aschwanden \& Aschwanden (2006),  
and we cannot rule out the possibility that the radiation losses are 
significant compared to the rate of change of thermal energy.  

It is also important to compare the transition region luminosity with the power carried
out of the X-ray emitting gas by thermal conduction.  This power is deposited in cooler
gas, where it must either be radiated away at transition region or lower temperatures,
or else heat the gas to flare temperatures (evaporation, or dE/dt of the X-ray plasma).
We estimate the conductive power as

\begin{equation}
P_{cond} ~=~ \kappa T^{7/2} A_{cross} / L_{1/2}~ = ~ 2 \kappa T^{7/2} g f^\prime A^{1/2} 
\end{equation}

\noindent
where $\kappa$ is the Spitzer conduction coefficient, $A_{cross}$ and  
$L_{1/2}$ are the cross sectional area and half length of the loops. The factor g relates
$ A_{cross} / L_{1/2}$ to the quantity $A^{1/2}$ that we can derive from the area of
the X-ray emission measurable from the RHESSI images.  It scales approximately as $(D/L)^{3/2}$,
where D is the loop diameter and L the loop length, and it depends on the angle from which
the loops are viewed.  For values L $\sim$ 3 to 10 D, g varies from 0.02 to 0.15.  The values
of $P_{cond}$ in Table 3 assume g = 1 to avoid introducing a subjective correction factor, so
we expect that they are overestimated by about an order of magnitude.  
The factor $f^\prime$ is the areal filling factor
of the flare loops at the chromosphere, $f^\prime~\le~1$.  Again, we tabulate values for
$f^\prime~=~1$ in Table 3 because $f^\prime$ is unknown.  Table 3 compares the 
estimated conductive powers with radiative losses and the rate of increase of thermal
energy in the X-ray emitting plasma.  The predicted conductive powers greatly
exceed the sum $L_{TR} + dE/dt$.  This implies either that the conductive
power is a serious overestimate or that other radiative losses are important.

The other likely important radiative loss that could balance the conductive power
is \LA.  We can obtain an upper limit to the \LA losses from the \LB associated
with the flare.  In most cases no impulsive \LB is detected, but a
\LB enhancement in the Nov. 4, 2003 event is clearly seen at a level of 0.018
the O VI enhancement.  The radiative components of \LA and \LB are typically
in a ratio of 600:1 (e.g. Raymond et al. 2003), so the flare emission in
\LA could be as much as 10 times the O VI emission.  It is therefore comparable to
$L_{TR}$ and far smaller than $P_{cond}$.   

It is also possible that thermal conduction could carry the energy to still lower
temperatures, and that chromospheric emission at optical and UV wavelengths balances
$P_{cond}$.  However, Machado et al. (1990) find that all the other chromospheric emission
in a flare adds to about as much emission as \LA.  Moreover,
thermal conduction is ineffective at such
low temperatures, and for values of $P_{cond}$ and A from Tables 2 and 3 temperature
gradients of order 100 K/m at $2 \times 10^4$ K would be needed to carry the energy
below the \LA emission zone, making the chromosphere only 0.1 km thick.  
Strong UV and optical emission are observed with TRACE and SORCE during the impulsive
phases of flares (e.g. Emslie et al. 2005), but this emission could result from direct
heating by particle beams.  It could also be produced by absorption of the half of the flare
X-ray emission that illuminates the solar surface, but that reprocessed radiation would
not help balance thermal conduction losses from the flare loops.  The short duration
of the optical and UV emission, the enormous temperature gradients required for 
thermal conduction to carry the energy, and the models that indicate that 
chromospheric emission is comparable to \LA emission (Machado et al. 1980) lead to the
conclusion that the optical/UV emission is mostly generated by particle beams impinging
on the lower chromosphere and X-ray illumination.   Thus the transition region
and chromospheric emission are smaller than dE/dt except when dE/dt drops at the
end of the impulsive phase, and radiative losses are unlikely to ever be comparable to the
conductive losses shown in Table 3.

We conclude the Equation 5 greatly overestimates the conductive
losses from the X-ray emitting plasma.  The classical conduction formula gives
results 2 to 10 times smaller than saturated thermal conduction (Balbus \&
McKee 1982) for the temperatures and size scales in Table 2 and densities
obtained from the emission measure and volume, $(EM/V)^{1/2}$, so saturation
of the conduction does not resolve the discrepancy.  The factor g is
unlikely to account for the difference, because the estimates of $P_{cond}$
exceed the radiative loss and evaporative heating terms by 
two orders of magnitude in many cases.  Therefore, the $P_{cond}$ estimate is 
too large because $f^\prime < < 1$.  That is, the area
of the flare flux tubes at transition region and chromospheric heights is
much smaller than the area measured from the RHESSI images.  Schrijver
et al. (2006) observed a flare with TRACE and other instruments.  The 
195 \AA\/ and 1600 \AA\/ images showed the footpoints of the loops, and the typical 
width was $1.4 \times 10^8$ cm, as opposed to length scales $A^{1/2}$ of order 
$1.5 \times 10^9$ cm estimated from the RHESSI areas. Presumably, the magnetic
flux tubes narrow down toward lower heights, reducing the conductive power
by an order of magnitude or more.

We can also compare the $L_{TR}$ with the rate of thermal 
energy increase in the X-ray emitting plasma and the rate of deposition of
non-thermal energy.  The luminosity at transition region temperatures is

\begin{equation}
L_{TR} ~=~ n^2 P A f^\prime h
\end{equation}

\noindent
where $n$ is the density, P is the average radiative loss rate coefficient at transition region temperatures, 
about $4 \times 10^{-22}~\rm erg~cm^3~s^{-1}$, and $h$ is the thickness of the transition region.  The
rate of thermal energy input is

\begin{equation}
\frac{dE_{ev}}{dt} ~=~ \frac{3 k T}{\mu}  \dot{m} ~=~ 2.4 \times 10^{15}~\frac{T}{1~keV} ~\dot{m}
\end{equation}

\noindent
where $k$ is the Boltzmann constant, $\mu$ the effective atomic weight, $\dot{m}$ is the rate of
mass evaporation into the flare loops, $T$ is the temperature of the X-ray loops and
$dE_{ev} / dT$ is the increase in thermal energy due to evaporation, as opposed to the total increase
due to heating of coronal plasma.  Since 

\begin{equation}
\dot{m} ~=~ \mu n A f^\prime v
\end{equation}

\noindent
for a flow speed $v$ through the transition region,

\begin{equation}
L_{TR} ~=~ \frac{n}{10^{13}}~\frac{1~keV}{T}~\frac{h}{v}~\frac{dE_{ev}}{dt}
\end{equation}

\noindent
For $L_{TR}\sim 0.1 dE/dt$ as in Table 3 and velocities of order 100 $\rm km~s^{-1}$ as
inferred from Figure 3, a region of 100 km thickness would imply a density near
$10^{12}~\rm cm^{-3}$.

The O VI luminosity could be generated with a conventional, but high pressure, transition
region.  For instance, a region with a density of $10^{12}~\rm cm^{-3}$ and thickness 10 km
could match the typical areas and O VI luminosities in Table 2.  However, much higher
densities have been suggested for the temperature range where O VI forms.
For the 2002 July 23 flare, Share et al. (2004) 
found densities of $10^{14}$ to $10^{15}~\rm cm^{-3}$ at
temperatures of $2 \times 10^5$ to $7 \times 10^5$ K based on
the $3 \gamma / 2 \gamma$ ratio of positron annihilation.  For
the 2003 Nov. 2 flare, the RHESSI data are consistent with
annihilation at chromospheric temperatures and densities, or
with emission from warm, dense material at $5~-~10 \times 10^4$ K (G. Share
2006, private communication).  Doschek et al. (1977) 
obtained a density of $10^{13}~\rm cm^{-3}$ from transition
region lines during an M-class flare, so $10^{14}~\rm cm^{-3}$ seems
plausible for these X-class flares.  For the lower end of the
density range given by Share et al.  and $v$ = 90 km~s$^{-1}$
as derived for the O~VI line, the values $L_{TR} / \dot{E}$ implied
by Table 2 yield values of $h$ of order 0.1 km.  We note that equation
9 allows larger values of h if most of dE/dt is attributed to heating
of coronal gas rather than evaporation.  However, one would predict
very small values of h from the emission measures and areas in Table 2
for the density range proposed by Share et al. (2004). 

Such a small value of $h$ seems difficult to explain with Coulomb
collisions as the only mechanism that stops the beam of energetic
particles impinging on the chromosphere.  It seems possible that
sudden deposition of energy might 
produce plasma turbulence, and this turbulence could help to
stop the particles, further increasing the energy deposition rate
in a positive feedback loop to generate heating in a very thin layer.
In any case, the O VI emission detected by UVCS places a firm upper
limit on the emission measure at the densities and temperatures inferred
from the RHESSI observation of the 23 July, 2002 event.

Finally, for the 2003 July 23 event, Table~3 shows that the minimum
energy deposition by non-thermal particles, $L_{NT}$, is 2 to 15 times dE/dt
and another order of magnitude larger than $L_{TR}$.  Chromospheric
losses, which are a few times $L_{TR}$, combined with dE/dt can
balance $L_{NT}$ during several of the UVCS exposures, but fail 
by an order of magnitude during the exposure beginning at 00:24 UT.
Models of the flaring chromosphere by Allred et al. (2005) show 
that the He II $\lambda$ 304 emission and the continuum emission 
can substantially exceed the \LA emission, particularly if the flux 
in the non-thermal beam is large, and this might account for the
apparent difference between sources and sinks of non-thermal energy
during this time period.

\section{Summary}

We have determined the radiative losses at transition region temperatures for the
impulsive phases of five X-class solar flares based on UVCS measurements of O VI photons
scattered from coronal O VI ions, and we have compared them with X-ray 
observations.  $L_{TR}$ exceeds
the X-ray luminosities early in the impulsive phase, but it is small compared to the
rate of thermal energy increase in the flare loops, dE/dt, unless the filling factor $f$ 
is $\sim$ 0.01, at the low end of the plausible range.  Estimates of
the chromospheric radiation and \LA indicate that they are
comparable to $L_{TR}$.  We note that our
comparison applies only to the impulsive phase of the flare and to the
efficiency with which plasma is heated to X-ray emitting temperatures.  On
longer time scales, the hot plasma cools by both local emission and by
conductive transport to lower temperatures, where it is radiated away in the EUV.

In the case of the
2002 July 23 flare, dE/dt is a significant fraction of the estimated minimum
rate of energy deposition by non-thermal particles.  Estimates of thermal
conduction power exceed the sum of transition region and chromospheric luminosities 
plus dE/dt, indicating that the simple estimate of conductive power is 
an overestimate.  This is probably due to a small filling factor of flux
tube area at the solar surface.  The ratios of $L_{TR}$ to the 
rates of thermal energy increase indicate that the emitting region is
about 100 km thick for densities of order $10^{12}~\rm cm^{-3}$, though 
a much smaller thickness would be implied by the high densities inferred
from RHESSI observations of the positron annihilation line. 

\bigskip
This work was supported by NASA grants NAG5-12827 and NNG06GI88G to the Smithsonian Institution.
It benefited from the workshop on Coronal Mass Ejections held at Elmau and at ISSI and
from the SHINE workshop in Big Sky, Montana.  We are grateful to the observers who
issue Max Millenium major flare watches, which helped us to obtain the UVCS observations.  We thank
Gerry Share for many helpful comments.  GDH 
thanks Brian Dennis and Leah Haga for help in the analysis of the X-ray data and 
acknowledges partial support from the RHESSI Project.  We thank an anonymous referee
for helpful comments.


\newpage

\begin{figure}
\plottwo{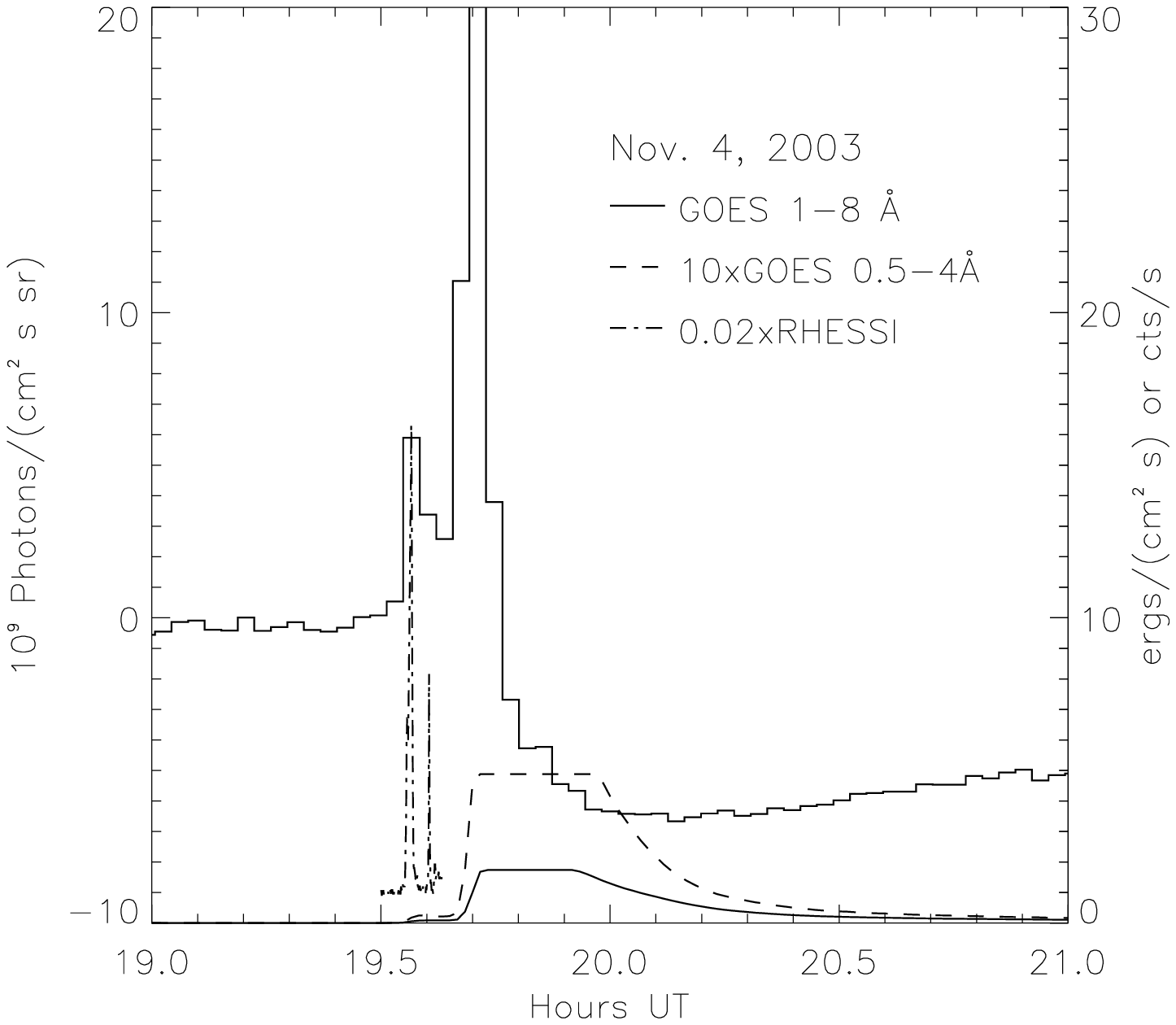}{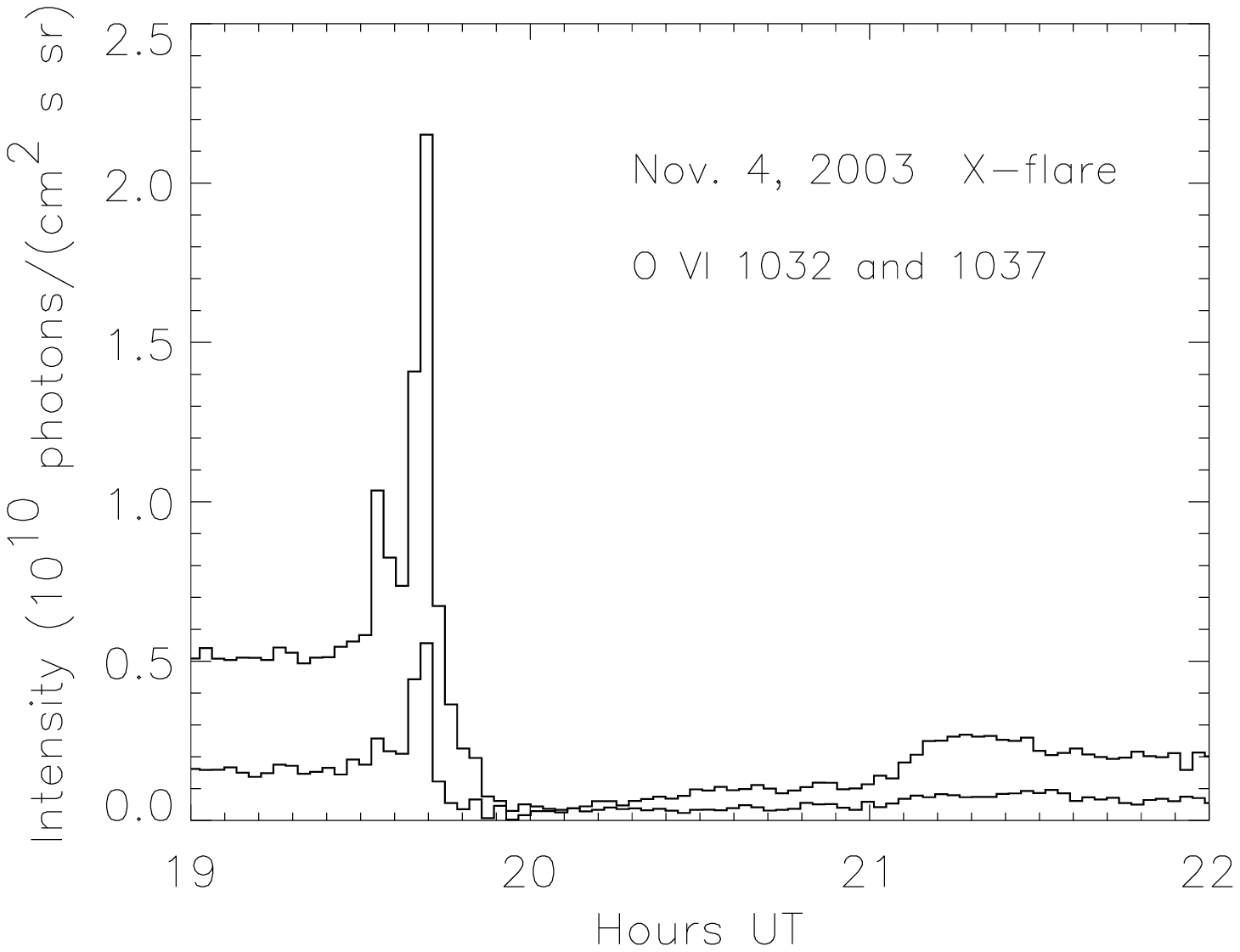}
\caption{ Left panel shows the O VI intensity (histogram) and X-ray fluxes for 
the Nov. 4, 2003 event. The pre-flare O VI intensity has been subtracted.
The continuous curves show the 1-8 \AA\/ (solid) and 0.5-4 (dashed) GOES X-ray
fluxes (right hand Y axis scale in $\rm erg~cm^{-2}~s^{-1}$) and RHESSI count 
rate (dot-dash) for photons above 50 keV (right hand axis scale in $\rm cts~s^{-1}$).
The flat portions of the GOES X-ray fluxes are due to saturation.
Right panel shows intensities of the O VI $\lambda \lambda$ 1032 and 1037
lines as functions of time, showing the increase in the intensity ratio to 
approximately 4:1 during the impulsive brightening.
\label{fig1}}
\end{figure}

\newpage

\begin{figure}
\epsscale{0.9}
\plotone{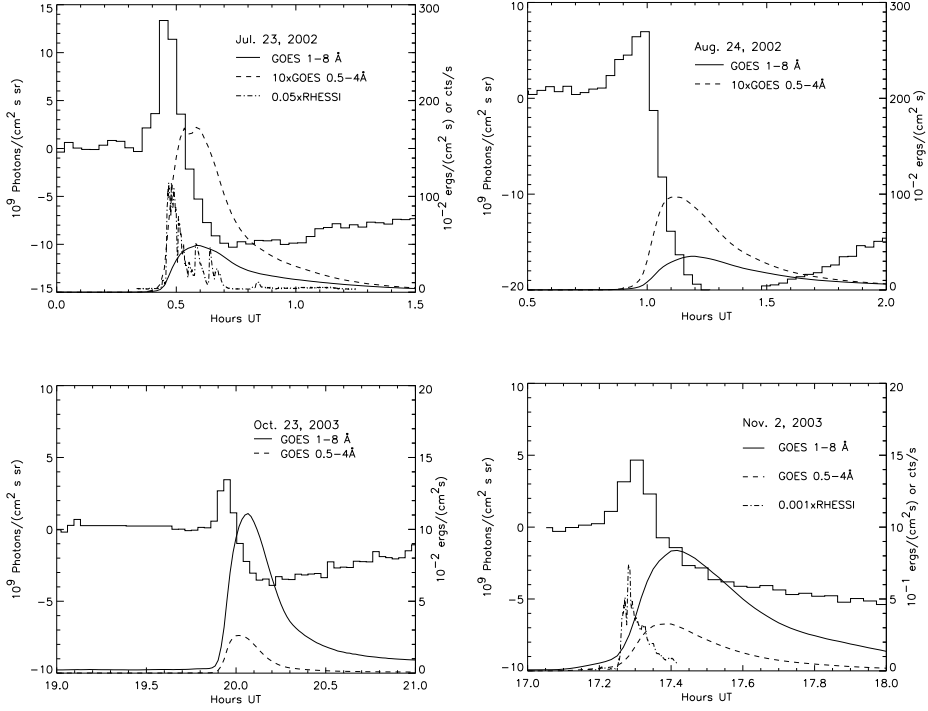}
\caption{
Figure 2.  Time histories of the O VI $\lambda$1032 intensities of 4 of the events
observed by UVCS (histogram).  The pre-flare O VI intensity has been subtracted.
The continuous curves show the 1-8 \AA\/ (solid)
and 0.5-4 (dashed) GOES X-ray fluxes (right hand Y axis scale in $\rm erg~cm^{-2}~s^{-1}$)
and RHESSI count rate (dot-dash) for photons above 50 keV (right hand axis scale in $\rm cts~s^{-1}$).
\label{fig2}}
\end{figure}
 
\newpage

\begin{figure}
\plotone{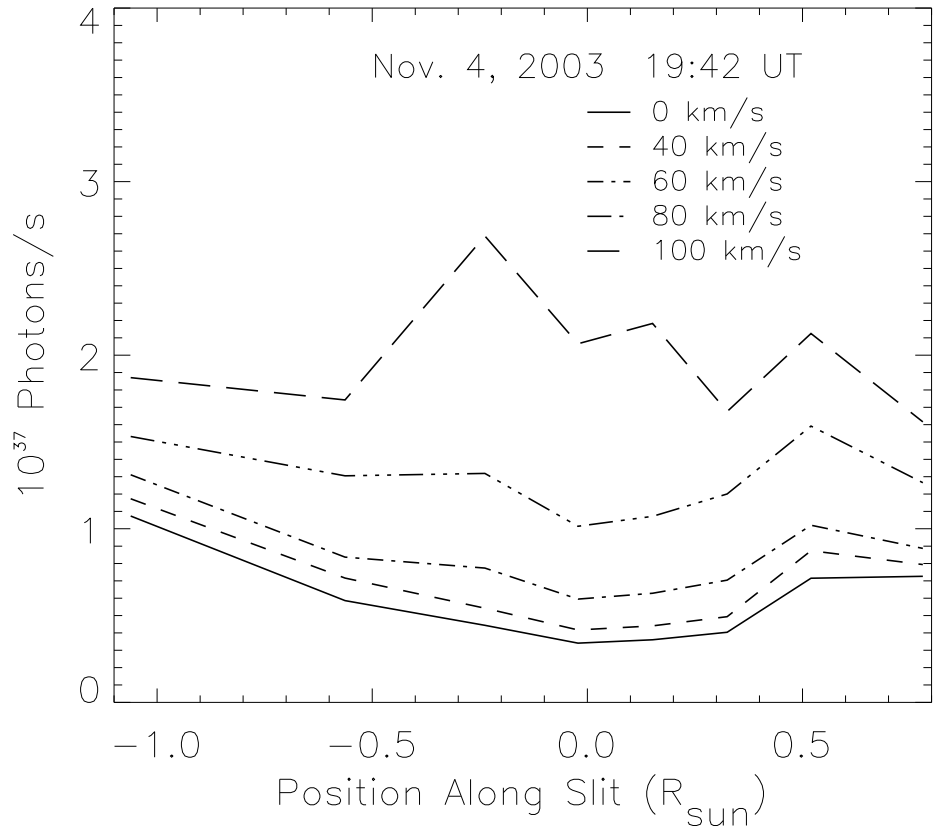}
\caption{
Derived O VI luminosities for different positions along the
slit for the exposure beginning at 19:42 UT during the Nov. 4, 2003 event.
\label{fig3}}
\end{figure}

\clearpage



\begin{deluxetable}{ lrrcc }
\tablecaption{UVCS Solar Flare Observations}
\tablewidth{0pt}
\tablehead{
\colhead{Event} &
\colhead{Flare Class} &
\colhead{Flare Position} &
\colhead{UVCS PA } &
\colhead{Height \RSUN\/ } 
}
\startdata
Jul. 23 2002 & X4.8 & S13E72 & 96$^\circ$  & 1.63  \\
Aug. 24 2002 & X3.1 & S02W81 & 260$^\circ$ & 1.63  \\
Oct. 23 2003 & X1.1 & S17E84 & 106$^\circ$ & 1.63  \\
Nov.  2 2003 & X8.3 & S14W56 & 245$^\circ$ & 1.63  \\
Nov.  4 2003 & $>$X28 & S19W83 & 262$^\circ$ & 1.63  \\


\enddata

\end{deluxetable}


\newpage

\normalsize

\begin{table}
\begin{center}
\centerline{Table 2}

\vspace*{2mm}
\centerline{Impulsive Phase O VI and X-ray Parameters}
\centerline{($10^{26}~erg~s^{-1}$, $10^{48}~cm^{-3}$,  $10^6$ K, $10^{18}~\rm cm^2$)}

\vspace{4mm}
\begin{tabular}{| lcrrrrrr |}
\hline \hline
 Event & $\rm T_{start}$ &  $L_{1032}$ & $EM_{UV}$ & $\rm L_{TR}$ & $T_6$ & $EM_{X}$ & $A_{18}$ \\
\hline

             &      &      &      &      &     &       &     \\
Jul. 23 2002 &      &      &      &      &     &       &     \\
 &  00:21:51 &  .19  &  .26   &  2.7   &  16.0 & 1.5   & 2.9 \\
 &  00:24:02 &  .28  &  .39   &  4.0   &  17.5 & 5.8   & 2.8 \\
 &  00:26:12 & 1.12  & 1.58   & 16.3   &  20.2 & 27.1  & 2.3 \\
 &  00:28:21 &  .95  & 1.33   & 13.8   &  23.8 & 100.2 & 1.9 \\
 &  00:30:32 &  .51  &  .71   &  7.4   &  22.1 & 171.2 & 2.3 \\
 &  00:32:42 &  .28  &  .40   &  4.1   &  20.4 & 209.4 & 2.7 \\

Aug. 24 2002 &      &      &      &      &      &      &   \\
 &  00:48:54 &  .03  &  .04  &   .43  &  12.4 & 0.8  & 26.0 \\
 &  00:51:03 &  .06  &  .09  &   .89  &  14.6 & 1.5  & 18.1 \\
 &  00:53:15 &  .13  &  .18  &  1.89  &  16.3 & 2.9  & 11.6 \\
 &  00:55:24 &  .14  &  .20  &  2.03  &  17.6 & 6.3  &  8.9 \\
 &  00:57:33 &  .33  &  .46  &  4.73  &  19.8 & 14.4  &  -   \\
 &  00:59:44 &  .47  &  .66  &  6.82  &  22.2 & 37.5  &  -   \\   

 Oct. 23 2003 &      &      &      &      &      &      &    \\
  &  19:52:16 &  .03  &  .04   &   .40   &  16.2 & 1.3   &  -  \\
  &  19:54:27 &  .16  &  .23   &  2.33   &  18.8 & 7.7   &  -  \\
  &  19:56:37 &  .20  &  .29   &  2.96   &  19.7 & 22.9   &  -  \\
  &  19:58:46 &  .18  &  .26   &  2.68   &  18.6 & 39.0   & 0.9 \\

 Nov.  2 2003 &      &      &      &      &      &      &  \\
  &  17:12:37 &  .01  &   .01  &   .10  &  20.6 & 27.3  & 1.9 \\
  &  17:14:48 &  .23  &  .32   &  3.27  &  21.6 & 51.2  &10.9  \\
  &  17:16:58 &  .43  &  .60   &  6.25  &  24.9 & 113.7  & 3.5 \\
  &  17:19:07 &  .16  &  .22   &  2.29  &  25.8 & 210.6  & 2.6 \\

  Nov.  4 2003 &      &      &      &      &     &       &       \\
   &  19:31:12 &  .19  &  .27   &  2.7    &  15.5 & 1.6   & 1.5 \\
   &  19:33:22 & 1.27  & 1.78   & 18.5    &  18.9 & 22.0   & 2.1 \\
   &  19:35:32 &  .84  & 1.18   & 12.2    &  17.9 & 39.5   & 1.0 \\
   &  19:37:45 &  .70  &  .98   & 10.1    &  16.9 & 42.1   &  -  \\
   &  19:39:56 & 2.71  & 3.80   & 39.4    &  22.9 & 95.9  &  -  \\
   &  19:42:06 & 5.32  & 7.47   & 77.5    &   -   & -     &  -  \\
   &  19:44:15 & 1.75  & 2.46   & 25.5    &    -  & -     &  -  \\
   &           &       &        &         &       &        &     \\

\hline
\end{tabular}
\end{center}
\end{table}


\normalsize

\begin{table}
\begin{center}
\centerline{Table 3}

\vspace*{2mm}
\centerline{Radiative Losses, Thermal, Non-thermal and Conductive }

\centerline{Powers ($10^{26}~erg~s^{-1}$)}

\vspace{4mm}
\begin{tabular}{| lrrrrrr|}
\hline \hline
Event & $\rm T_{start}$ & $L_{TR}$ & $L_{therm}$ & $P_{cond}$ & dE/dt & $L_{NT}$ \\
\hline
             &      &      &      &    &      &    \\
Jul. 23 2002 &      &      &      &    &      &    \\
 &  00:21:51 & 2.7 & 0.7 & 1140~& 38.~ & 156 \\
 &  00:24:02 & 4.0 & 2.3 & 1550~& 72.~ & 1240 \\
 &  00:26:12 & 16.0& 9.0 & 2330~& 228.~ & 415 \\
 &  00:28:21 & 13.8& 29.8& 3750~& 161.~ & 318 \\
 &  00:30:32 & 7.4 & 54.2& 3180~& 49.~ & 161 \\
 &  00:32:42 & 4.1 & 69.3& 2610~& -2.~ & 178 \\

Aug. 24 2002 &      &      &      &    &      &   \\
 & 00:48:54 & 0.4  & 0.5  & 1430~& 39.~ &  \\
 & 00:51:03 & 0.9  & 0.8  & 2090~& 81.~ &  \\
 & 00:53:15 & 1.9  & 1.3  & 2470~& 77.~ &  \\
 & 00:55:24 & 2.0  & 2.5  & 2830~& 116.~ &  \\

Oct. 23 2003  &      &      &      &    &      &  \\
 & 19:58:46 & 2.7 & 14.7& 1110~& 9.~   &  \\

Nov.  2 2003 &      &      &      &   &      &   \\
 &  17:12:37 & 0.1 & 9.0 & 2300~ &  41.~ &  \\
 &  17:14:48 & 3.3 & 16.2& 6400~ & 514.~ &  \\
 &  17:16:58 & 6.2 & 32.6& 6040~ & 347.~ &  \\
 &  17:19:07 & 2.3 & 59.4& 5810~ & 143.~ &  \\

Nov.  4 2003 &      &      &      &     &      & \\
 &  19:31:12 & 2.7  & 0.7  &  749~&  32.~ &  \\
 &  19:33:22 & 18.5 & 7.7  & 1760~& 155.~ &  \\
 &  19:35:32 & 12.2 & 15.6 & 1015~&  -6.~ &  \\
             &     &      &      &      &      &  \\

\hline
\end{tabular}
\end{center}
\end{table}

%


\end{document}